\documentclass[11pt,a4paper]{article}

\usepackage[psamsfonts]{amssymb}
\usepackage{amsmath}

\author{H. Mohseni Sadjadi\footnote{mohsenisad@ut.ac.ir}
\\ {\small Department of Physics, University of Tehran,}
\\ {\small P. O. B. 14395-547, Tehran 14399-55961, Iran}}
\title{Crossing the phantom divide line in the Holographic dark energy model in a closed universe}

\begin{document}
\maketitle \abstract{Conditions needed to cross the phantom divide
line in an interacting holographic dark energy model in closed
Friedmann- Robertson-Walker universe are discussed. The probable
relationship between this crossing and the coincidence problem is
studied.}

\section{Introduction}
One of the candidates proposed to explain the present acceleration of the universe \cite{acc}, is the
dark energy model which assumes that nearly $\%70$ of the universe is filled of an exotic
energy component with negative pressure. Based on observations, the density of (dark) matter and dark energy component must be of the same order today (known as coincidence problem)\cite{coin}. Also, based on recent data, the dark energy component seems to have an equation of state parameter $w<-1$ at the present epoch, while $w>-1$ in the past \cite{cross}. One way to explain these data, is to consider dynamical dark energy with proper interaction with matter \cite{inter}.

In \cite{coh}, it was found that formation of black holes requires a relationship between ultraviolet and infrared cutoffs. In this context the total energy, $E$, in a region of size $L$, must be less than (or equal to) the energy of a black hole of the same size, i.e.,  $E\leq L M_p^2$, where $M_p$ is the Planck mass. In terms of energy density, $\rho$, this inequality can be rewritten as $\rho \leq M_p^2L^{-2}$. Based on this result, in \cite{Li}, an expression for a dynamical dark energy (dubbed as the holographic dark energy) was proposed : $\rho_d=3c^2M_p^2L^{-2}$, where $c$ is a numerical constant. Different choices may be adopted for the infrared cutoff of the universe, e.g., particle horizon, Hubble horizon, future event horizon and so on \cite{holo}. In a noninteracting model, if we take the particle horizon as the infrared cutoff, we are unable to explain the accelerated expansion of the universe \cite{Li}. Besides an appropriate equation of state parameter for dark energy or dark matter cannot be derived if one chooses the Hubble horizon as the cutoff \cite{Hsu}. Instead, if we choose the future event horizon, although the present accelerated expansion of the universe may be explained \cite{Li}, but the coincidence problem still unsolved. This problem can be alleviated by considering suitable interaction between dark matter and holographic dark energy.

In this paper we consider a closed Friedmann- Robertson- Walker (FRW) universe (we don't restrict ourselves to small spatial curvature limit) and assume that the universe is composed of two interacting perfect fluids: holographic dark energy and cold (dark) matter. A general (as far as possible) interaction between these components is considered. We allow the infrared cutoff to lie between future and particle event horizons. After some general remarks about the properties of the model, we discuss the conditions needed to cross the phantom divide line (transition from quintessence to phantom phase). We show that this crossing poses some conditions on parameters of the model and using an example, we show that this can alleviate the coincidence problem (at least) at transition epoch.

We use units $\hbar=G=k_{B}=c=1$ throughout the paper.

\section{General properties of the model}
The FRW metric,
\begin{equation}\label{1}
ds^2=-dt^2+a^2(t)\left({dr^2\over {1-r^2}}+r^2(d\theta^2+\sin^2\theta d\phi^2)\right),
\end{equation}
describes a homogeneous and isotropic closed space time with scale factor $a(t)$.
We assume that this universe is filled with perfect fluids  and its energy momentum tensor is given by
\begin{equation}\label{2}
T_{\mu \nu}=(P+\rho)U_\mu U_\nu+Pg_{\mu \nu},
\end{equation}
where $U^\mu=(1,0,0,0)$, is the normalized four velocity of the fluids in the comoving coordinates, and $\rho$ and $P$ are energy density and pressure of the total fluid respectively.
Using Einstein's equation, one can obtain Friedmann equations
\begin{eqnarray}\label{3}
H^2&=&{8\pi\over 3}\sum_i \rho_i-{1\over a^2(t)}\nonumber \\
\dot{H}&=&-4\pi\sum_i(P_i+\rho_i)+{1\over a^2(t)}.
\end{eqnarray}
The subscript $i$  stands for the $ith$ perfect fluid and $H={\dot{a(t)}\over a(t)}$ is the Hubble parameter.  In this paper the universe is assumed to be composed of
dark energy component with pressure $P_d$ and energy density $\rho_d$,  and the cold (dark) matter whose  energy density is $\rho_m$. Although these components, due to their interaction, are not conserved
\begin{eqnarray}\label{4}
\dot{\rho_d}+3H(\rho_d+P_d)&=&-Q\nonumber \\
\dot{\rho_m}+3H\rho_m&=&Q,
\end{eqnarray}
but vanishing of covariant divergence of the energy momentum tensor (\ref{2}), yields the conservation equation
\begin{equation}\label{5}
\dot{\rho}+3H(P+\rho)=0,
\end{equation}
for the whole system. In this two-component universe the Friedmann equations reduce to
\begin{eqnarray}\label{6}
H^2&=&{8\pi \over 3}(\rho_m+\rho_d)-{1\over a^2(t)}\nonumber \\
\dot{H}&=&-4\pi(P_d+\rho_d+\rho_m)+{1\over a^2(t)}.
\end{eqnarray}
The first equation can be rewritten as
\begin{equation}\label{7}
\Omega_m+\Omega_d=1+\Omega_k,
\end{equation}
where $\Omega_m={\rho_m\over \rho_c}$,
$\Omega_d={\rho_d\over \rho_c}$ and the geometrical parameter is defined through  $\Omega_k={1\over a^2(t)H^2}$.
 The critical energy density  $\rho_c$ is defined by $\rho_c={3H^2\over 8\pi}$. Note that for a flat universe, $\rho=\rho_c$ and $\Omega_m+\Omega_d=1$.  Different models have been proposed for the dark energy component of the universe. Here we adopt holographic dark energy, which in terms of the infrared cutoff of the universe, $L$, can be expressed as
\begin{equation}\label{8}
\rho_d={3c^2\over 8\pi L^2}.
\end{equation}
In \cite{hua} the infrared cutoff was chosen as $L_f=a(t)\sin y_f$ where
$y_f$ is
\begin{eqnarray}\label{9}
y_f&=&\int_t^\infty {dt\over a(t)} \nonumber \\
&=&\int_0^{r_f}{dr\over {\sqrt{1-r^2}}}.
\end{eqnarray}
In this way $L_f$ is the radius of future event horizon measured on the sphere of the horizon \cite{hua}.
In the presence of bigrip \cite{bigrip} at $t=t_s$, $\infty$ in (\ref{9}) must be replaced with $t=t_s$. In the flat case this cutoff reduces to
$L_f=R_h=a(t)\int_t^\infty {dt\over a(t)}$. A similar choice is to
take an infrared cutoff based on the particle horizon, i. e.,
$L_p=a(t)\sin y_p$, where
\begin{eqnarray}\label{10}
y_p&=&\int_0^t {dt\over a(t)} \nonumber \\
&=&\int_0^{r_p}{dr\over {\sqrt{1-r^2}}}.
\end{eqnarray}
As proposed in \cite{odin}, in general, $L$ can be taken as a combination of both $L_p$ and $L_f$. In this paper we take the cutoff as:
\begin{equation}\label{11}
L=\alpha L_p+\beta L_f; \,\,\, 0\leq\alpha\leq1, \,\, 0\leq\beta\leq 1.
\end{equation}
For $\alpha=0, \beta=1$, we get $L=L_f$ and $L=L_p$ is obtained when $\beta=0, \alpha=1$.
If we attribute an entropy, $S$,  to the surface $A=4\pi L^2$:
\begin{equation}\label{12}
S={A\over 4}=\pi L^2,
\end{equation}
then the thermodynamics second law implies  $\dot{L}>0$.
Using
\begin{equation}\label{13}
\dot{L}=HL-\beta\cos y_f+\alpha\cos y_p,
\end{equation}
we find that the second law of thermodynamics is satisfied when
\begin{equation}\label{14}
X<1,
\end{equation}
where
\begin{equation}\label{15}
X={{\beta\cos y_f-\alpha\cos y_p}\over c}\Omega_d^{1\over 2}.
\end{equation}
Note that
\begin{equation}\label{16}
y_p+y_f=\int_0^\infty {dt\over a(t)}:=\gamma,
\end{equation}
is a functional of $a(t)$ and is time independent:
$\dot{\gamma}=0$. In the limit $\Omega_k\ll \Omega_d$,
\begin{equation}\label{17}
\alpha \sin y_p+\beta \sin y_f={L\over a(t)}={c({\Omega_k\over
\Omega_d})^{1\over 2}}
\end{equation}
implies: $y_f,y_p \ll 1$.

The time evolution of the geometrical parameter is obtained as
\begin{equation}\label{18}
\dot{\Omega_k}=-2H\Omega_k\left(1+{\dot{H}\over H^2}\right),
\end{equation}
and, the ratio
$r={\Omega_m\over \Omega_d}$ satisfies
\begin{equation}\label{19}
\dot{r}=(r+1)\left({Q\over
\rho_d}+3H\omega r\right).
\end{equation}
Note that $r$ determines the ratio of (dark) matter to dark energy density and is of order unity in the present epoch. One can also study the time evolution of  $\mathcal{K}={\Omega_k\over \Omega_d}$.  Using $\mathcal{K}=({L\over ca(t)})^2$, it is straightforward to show that in order that $\dot{\mathcal{K}} \lessgtr 0$, we must have
\begin{equation}\label{20}
(\alpha
\cos \gamma-\beta)\cos y_f+\alpha\sin \gamma\sin y_f\lessgtr 0.
\end{equation}
Therefore the behavior of this ratio with respect to the comoving time depends on the parameters of the model, e.g.  for $\alpha=0, \beta=1$ (i.e., when the future event horizon is considered), it is decreasing.

\section{Crossing the phantom divide line and the coincidence problem}
In this part we study the ability of our model to describe $w=-1$ crossing (crossing the phantom divide line)
and the probable relationship between this event and the lower bound of $r$ (coincidence problem).

By using
$HL=c\Omega_d^{-{1\over 2}}$, we arrive at
\begin{equation}\label{21}
{\dot{H}\over H^2}+{c\over 2H^2L}\dot{\Omega_d}{\Omega_d^{-{3\over
2}}}={{\beta\cos y_f- \alpha\cos y_p}\over HL}-1.
\end{equation}
Hence from
\begin{equation}\label{22}
w=-1-{2\over 3}{{{\dot{H}\over H^2}-\Omega_k}\over{1+\Omega_k}},
\end{equation}
we find out
\begin{equation}\label{23}
w=-{1\over 3}-{2X\over
3(1+\Omega_k)}+{1\over {3H(1+\Omega_k)}}
{\dot{\Omega_d}\over \Omega_d}.
\end{equation}
The universe is accelerating for  $\ddot{a}>0$, or equivalently when $w<-{1\over 3}$. In this case
\begin{equation}\label{24}
\dot{\Omega_d}< 2H\Omega_d X,
\end{equation}
which by considering the thermodynamics second law results in:
\begin{equation}\label{25}
\dot{\Omega_d}<2H\Omega_d,
\end{equation}
or equivalently, ${\Omega_d\over a(t)^2}$ is a decreasing function of comoving time.
$w$ can be also derived from the equation (\ref{19}), resulting
\begin{equation}\label{26}
w=-{\dot{\Omega_d}\over 3H\Omega_m}+{\dot{\Omega_k}\Omega_d\over 3H\Omega_m(1+\Omega_k)}-{Q\over 3H\rho_d}{\Omega_d\over \Omega_m}.
\end{equation}
Suppressing $\dot{\Omega_d}$ from (\ref{23}) and (\ref{26}) leads to
\begin{equation}\label{27}
w=-{\Omega_d\over 3(1+\Omega_k)}\left({Q\over H\rho_d}+1\right)-{2X\Omega_d\over 3(1+\Omega_k)}.
\end{equation}
Note that if $w=-1$ crossing is allowed, $\Omega_d$ must satisfy the equation
\begin{equation}\label{28}
\left(2X+{Q\over H\rho_d}+1\right)\Omega_d=3\left(1+\Omega_k\right),
\end{equation}
at transition time, i.e., this equation must have at least one root.
Moreover, to cross $w=-1$, $\dot{w}$ must be negative at $w=-1$. To determine $\dot{w}$, we note that
\begin{equation}\label{29}
\dot{X}=H\left(X^2+{1\over 2}\alpha(1+\Omega_k)X+\Omega_k\right),
\end{equation}
where  $\alpha=1+3w$. To derive the above equation we have used
\begin{equation}\label{30}
{1\over c}\Omega_d^{1\over 2}(-\beta \dot{y_f}\sin y_f+\alpha \dot{y_p}\sin y_p)=H\Omega_k.
\end{equation}
Substituting (\ref{29}) and (\ref{30}), in time derivative of (\ref{27}) results in
\begin{eqnarray}\label{31}
\dot{w}&=&
-{2H\Omega_d\over {1+\Omega_k}}[X^2+\left({\alpha\over 6}(3+\Omega_k)+{Q\over 3H\rho_d}+{1\over 3}\right)X+
{\Omega_k\over 3}\nonumber \\
&+&
{\alpha\over 6}(1+{Q\over H\rho_d})+{1\over 6H}({Q\over H\rho_d}\dot{)}].
\end{eqnarray}
In studying the divide line crossing, we intend to adopt the validity of thermodynamics second law. Hence it is more convenient to write (\ref{31}), at transition time, in the form
\begin{equation}\label{32}
\dot{w}=-{2H\Omega_d\over 1+\Omega_k}\left((X-1)(X-{1\over 3}(\Omega_k-{Q\over H\rho_d}-1))\right)-{\Omega_d\over{3(1+\Omega_k)}}\left({Q\over H\rho_D}\right)^..
\end{equation}
${\dot w}<0$ gives
\begin{equation}\label{33}
\left(X-1\right)\left(X-{1\over 3}\left(\Omega_k-{Q\over H\rho_d}-1\right)\right)>-{1\over 6H}\left({Q\over H\rho_d}\right)^..
\end{equation}

If the universe remains in the phantom phase after the
transition,  the cosmological evolution may be ended by a big rip singularity \cite{bigrip}. But, (\ref{28}), depending on the interaction $Q$, and the parameters $\alpha$ and $\beta$, may have more than one root for $\Omega_d$ when $w=-1$. This may allow another transition from phantom to quintessence phase and avoids bigrip singularity. In this case, in the transition from phantom to quintessence phase, we must have $\dot{w}>0$ at $w=-1$.

To go further we must specify the interaction term $Q$.  Note that $\rho_m={r\over r+1}\rho$ and $\rho_d={1\over r+1}\rho$. Hence if ${Q\over H\rho_d}$ is a function of $\rho_m$ and $\rho_d$, it  can be casted into the form $q(\rho,r)$, this may prompt us to assume, as a choice,  ${Q\over H\rho_d}=:q(\rho,r)$.
$X$ can be expressed in terms $r$ and $q$ as
\begin{equation}\label{34}
X={{3r-q}\over 2}+1.
\end{equation}
Time evolution of $q$ is given by the equation
\begin{eqnarray}\label{35}
\dot{q}&=&q_{,\rho} \dot{\rho}+q_{,r}\dot{r}\nonumber \\
&=& q_{,\rho}\dot{\rho}-Hq_{,r}(r+1)(\alpha+2X),
\end{eqnarray}
which at $w=-1$, reduces to
\begin{equation}\label{36}
\dot{q}=2Hq_{,r}(r+1)(1-X).
\end{equation}
To obtain the above equation we have used $\dot{r}=-H(1+r)(\alpha+2X)=-{1+\Omega_k\over \Omega_d}(\alpha+2X)H$,
and defined $q_{,x}={\partial q\over \partial x}$.
Collecting these results we obtain
\begin{equation}\label{37}
\dot{w}=-{2H\Omega_d \over 1+\Omega_k}\left((X-1)(X+{1\over 3}\left(1-\Omega_k+q-(r+1)q_{,r})\right)\right).
\end{equation}
$X<1$ implies $3r<q$, and
in order to cross the phantom divide line,
\begin{equation}\label{38}
X<{1\over 3}\left(\Omega_k+(r+1)q_{,r}-q-1\right)
\end{equation}
must be hold. As an example consider the interaction term as
$Q=\lambda_mH\rho_m+\lambda_dH\rho_d$, hence $q=\lambda_m r +\lambda_d$. Therefore
\begin{equation}\label{39}
\dot{w}=-{2H\Omega_d\over 1+\Omega_k}\left[(X-1)\left(X+{1\over 3}\left(1-\Omega_k+\lambda_d-\lambda_m\right)\right)\right].
\end{equation}
Here, the validity of thermodynamics second law implies
\begin{equation}\label{40}
(3-\lambda_m)r<\lambda_d,
\end{equation}
and the condition for crossing $w=-1$ is
\begin{equation}\label{41}
(3-\lambda_m)r<{\lambda_d+2\lambda_m+2\Omega_k-8\over 3}.
\end{equation}
So we have to assume
\begin{equation}\label{42}
(3-\lambda_m)r<Min.\{\lambda_d,{\lambda_d+2\lambda_m+2\Omega_k-8\over 3} \}.
\end{equation}
For $\lambda_m>3$, we have
\begin{equation}\label{43}
r>{Min.\{\lambda_d,{\lambda_d+2\lambda_m+2\Omega_k-8\over 3} \}\over 3-\lambda_m}.
\end{equation}
If $Min.\{\lambda_d,{\lambda_d+2\lambda_m+2\Omega_k-8\over 3} \}<0$, the above inequality poses a lower bound on $r$ at $w=-1$ and alleviates the coincidence problem. This lower bound depends on the interaction parameters and $\Omega_k$ at transition time. E. g., if we take $\Omega_k=0.02$ at transition time, all models satisfying $3<\lambda_m<3-{7\over 3} Min.\{\lambda_d,{\lambda_d+2\lambda_m-7.96\over 3} \}$ give rise
to $r>{3\over 7}$ in accordance with recent data \cite{coin}.

At the end it is worth to note that in the above example, for $X=1$, i.e. when the expansion is adiabatic $\dot{S}=0$, (\ref{29}), and (\ref{31}) imply that the higher time derivatives of $X$ and $\dot{w}$ are also zero at $w=-1$. In this situation, $w=-1$ is denoted as the point of infinite flatness and can occur only at $t\to \infty$. So $w=-1$ is not crossed in this case.
Instead, if we assume that ${Q\over \rho_d}=f(\rho,r)$, then  at $w=-1$, $({Q\over \rho_d H})^.=f_{,r}(r+1)(1-X)-\Omega_k f$ (see (\ref{22})).  If $X=1$, the sign of $f$ determines whether $w=-1$ is crossed or not. E.g., consider $Q=\lambda \rho_m \rho_d$ \cite{new}. In this model  $({Q\over H\rho_d})^.=-\lambda \rho_d[\Omega_k r+(X-1)]$
and we can have $X=1$, meanwhile $w=-1$ is crossed for $\lambda<0$.
\section{Conclusion}
Holographic dark energy model in a closed FRW universe
(but not necessarily with a small spatial curvature), was considered.
The infrared cutoff was taken to be lie between particle and future event horizons (see (\ref{11})),
and dark energy and dark matter were assumed to interact via a general interaction source (see (\ref{4})).
The condition of validity of thermodynamics second law for the infrared cutoff was obtained (see (\ref{14})).
Using equations derived for equation of state parameter of dark energy and its time derivative,
condition required for crossing the phantom divide line was derived (see (\ref{33})).
By adopting thermodynamics second law and restricting the interaction term to special forms,
this condition was reduced to a more compact form (see (\ref{38})), revealing the probable relationship
between $w=-1$ crossing and the coincidence problem (see(\ref{43})).  At the end we discussed
the possibility of adiabatic expansion at transition time.

As a consequence of our lack of knowledge about the nature of dark
energy and dark matter, the form of interaction term, $Q$, is
still unknown. Also different choices for the infrared cutoff are
used in the literature. The viability of a model, with a specific
$Q$ and a particular infrared cutoff, corresponds to its agreement
with astrophysical data. Using the general result (\ref{33}) one
can examine whether a proposed model (characterized by $Q$ and the
cutoff $L$ (defined by (\ref{11})) is compatible with phantom
divide line crossing and meanwhile satisfies thermodynamics second
law. However these phenomenological models, can give us some clues
to refine our view about the realistic dark energy model.

{\bf{Acknowledgment}}

The author would like to thank the University of Tehran for supporting him under the grant
provided by its Research Council. This work was partially supported by the ''center
of excellence in structure of matter'' of the Department of Physics.


\begin{thebibliography}{99}
\bibitem{acc}S. Perlmutter et al., Nature (London) \textbf{391}, 51 (1998);
A. G. Riess et al. (Supernova Search Team Collaboration), Astron.
J. \textbf{116}, 1009 (1998); Astron. J. \textbf{117}, 707 (1999);
S. Perlmutter et al. (Supernova Cosmology Project Collaboration),
Astrophys. J. \textbf{517}, 565 (1999).
\bibitem{coin}
P. J. Steinhardt, in Critical Problems in Physics, Eds. V.L.
Fitch, D. R. Marlow, and. M. A. E. Dementi, Princeton University
Press, 1997; N. Straumann, arXiv:astro-ph/0009386v1; Y. Fujii,
Phys. Rev. D \textbf{62}, 064004 (2000); L. P. Chimento, A. S.
Jakubi, and D. Pavon, arXiv:astro-ph/0010079v1; D. T. Valentini,
L. Amendola, Phys. Rev. D \textbf{65}, 063508 (2002);  V. Sahni,
Lect. NotesPhys. \textbf{653}, 141 (2004);  P. P. Avelino, Phys.
Lett. B \textbf{611}, 15 (2005); R. Curbelo, T. Gonzalez, and I.
Quiros, Class. Quant. Grav. \textbf{23}, 1585 (2006); S. Nojiri,
and S. D. Odintsov, Gen. Rel. Grav. \textbf{38}, 1285 (2006); L.
Amendola, S. Tsujikawa, and M. Sami, Phys. Lett. B \textbf{632},
155 (2006); B. Hu, and Y. Ling, Phys. Rev. D \textbf{73}, 123510
(2006); P. B. Almeida, and J. G. Pereira, Phys. Lett. B
\textbf{636}, 75 (2006); H. Wei, and R. G. Cai, Phys. Rev. D
\textbf{73}, 083002 (2006); S. Nojiri, and S. D. Odintsov, Phys.
Lett. B \textbf{637}, 139 (2006); J. Kujat, R. J. Scherrer, and A.
A. Sen, Phys. Rev. D \textbf{74}, 083501 (2006); H. M. Sadjadi,
and M. Alimohammadi, Phys. Rev. D \textbf{74}, 103007 (2006);  M.
Ishak, Found. Phys. \textbf{37}, 1470 (2007); J. Grande, J. Sola,
and H. Stefancic, J. Phys. A \textbf{40}, 6787 (2007); F. Melia,
arXiv:0711.4810v1 [astro-ph]; C. Gao, F. Wu,  X. Chen, and Y-G.
Shen , arXiv:0712.1394v4 [astro-ph]; H. Wei, and R. G. Cai, Phys.
Lett. B \textbf{663}, 1 (2008); M. Li, C. Lin, and Y. Wang, J.
Cosmol. Astropart. Phys. \textbf{05}, 023 (2008); K. Karwan, J.
Cosmol. Astropart. Phys. \textbf{05}, 011 (2008); J. H. He, and B.
Wang , arXiv:0801.4233v2 [astro-ph]; A. J. M. Medved ,
arXiv:0802.1753v2 [hep-th]; E. V. Linder, and R. J. Scherrer,
arXiv:0811.2797v1 [astro-ph]; M. Jamil, and F. Rahaman,
arXiv:0810.1444v2 [gr-qc]; S. d. Campo, R. Herrera, and D. Pavon,
Phys. Rev. D \textbf{78}, 021302 (2008); J. B. Jimenez, and A. L.
Maroto, arXiv:0812.1970v1 [astro-ph]; X. M. Chen, Y. Gong, and E.
N. Saridakis, arXiv:0812.1117v1 [gr-qc];  M. Quartin, M. O.
Calvao, S. E. Joras, R. R. R. Reis, and I. Waga , J. Cosmol.
Astropart. Phys. \textbf{05}, 007(2008).
\bibitem{cross}
U. Alam , V. Sahni, and A. A. Starobinsky, J. Cosmol. Astropart.
Phys. \textbf{06}, 008 (2004); V. K. Onemli and R. P. Woodard,
Phys. Rev. D \textbf{70}, 107301 (2004); B. Feng, X. Wang, and X.
Zhang, Phys. Lett. B \textbf{607}, 35 (2005); D. Huterer, and A.
Cooray, Phys. Rev. D \textbf{71}, 023506 (2005); S. Nojiri and S.
D. Odintsov, Phys. Rev. D \textbf{72}, 023003 (2005); S. Nesseris
and L. Perivolaropoulos, Phys. Rev. D \textbf{72}, 123519 (2005);
H. M. Sadjadi, and M. Alimohammadi, Phys. Rev. D \textbf{74},
043506 (2006); H. M. Sadjadi, and M. Honardoost, Phys. Lett. B
\textbf{647}, 231 (2007); M. Alimohammadi, and H. M. Sadjadi,
Phys. Lett. B \textbf{648}, 113 (2007); M. B. Lopez, and R.
Lazkoz, Phys. Lett. B \textbf{654}, 51 (2007); M. B. Lopez, and A.
Ferrera, J. Cosmol. Astropart. Phys. \textbf{10}, 011(2008); K.
Bamba, C. Q. Geng, S. Nojiri, and S. D. Odintsov,
arXiv:0810.4296v1 [hep-th]; K. Nozari, and M. Pourghasemi, J.
Cosmol. Astropart. Phys. \textbf{10}, 044 (2008); K. Nozari, N.
Behrouz, T. Azizi, and B. Fazlpour, arXiv:0808.0318v2 [gr-qc]; M.
B. Lopez, and P. V. Moniz, Phys. Rev. D \textbf{78}, 084019
(2008); L. N. Granda, and A. Oliveros, arXiv:0901.0561v2 [hep-th];
M. Alimohammadi, and A. Ghalee, Phys. Rev. D \textbf{79}, 063006
(2009).
\bibitem{inter}
L. Amendola, Phys. Rev. D \textbf{62}, 043511 (2000); W. Zimdahl,
D. Pavon, and L. P. Chimento, Phys. Lett. B \textbf{521}, 133
(2001);   G. Mangano, G. Miele, and V. Pettorino, Mod. Phys. Lett.
A \textbf{18}, 831 (2003); L. P. Chimento, A. S. Jakubi, D. Pavon,
and W. Zimdahl, Phys. Rev. D \textbf{67}, 083513 (2003); G. Farrar
and P. J. E. Peebles, Astrophys. J. \textbf{604}, 1 (2004); S. del
Campo, R. Herrera, and D. Pavon, Phys. Rev. D \textbf{70}, 043540
(2004); R. G. Cai, and A. Wang, J. Cosmol. Astropart. Phys.
\textbf{03}, 002 (2005); J. D. Barrow, and T. Clifton, Phys. Rev.
D \textbf{73}, 103520 (2006); S. Tsujikawa, and M. Sami, J.
Cosmol. Astropart. Phys. \textbf{01}, 006 (2007); E. J. Copeland,
M. Sami, and S. Tsujikawa, Int. J. Mod. Phys. D \textbf{15}, 1753
(2006); H. M. Sadjadi, J. Cosmol. Astropart. Phys. \textbf{02},
026 (2007); T. Clifton, and J. D. Barrow, Phys. Rev. D
\textbf{75}, 043515 (2007); L. P. Chimento, M. Forte, and G. M.
Kremer, arXiv:0711.2646v3 [astro-ph]; X. Fu, H. Yu, and P. Wu,
Phys. Rev. D \textbf{78}, 063001 (2008); M. Jamil, and M. A.
Rashidb, Eur. Phys. J. C  \textbf{58}, 111 (2008); G. C. Cabral,
R. Maartens, and  L. A. U. Lopez, arXiv:0812.1827v1 [gr-qc]; G.
Leon, arXiv:0812.1013v1 [gr-qc]; C. Feng, B. Wang, E. Abdalla, and
R. K. Su, arXiv:0804.0110v2 [astro-ph]; S. H. Pereira, J. F.
Jesus, arXiv:0811.0099v1 [astro-ph]; S. Feng Wu, P. M. Zhang, and
G. H. Yang, arXiv:0809.1503v3 [astro-ph]; X. Zhang
arXiv:0901.2262v1 [astro-ph]; C. G. Boehmer, G. C. Cabral, R.
Lazkoz, and R. Maartens , Phys. Rev. D \textbf{78}, 023505 (2008);
Y. Ma, Yan Gong, and X. Chen, arXiv:0901.1215v1 [astro-ph.CO]; S.
Chattopadhyay, and U. Debnath, arXiv:0901.2184v1 [gr-qc]; M. A.
Rashid, M. U. Farooq, and M. Jamil, arXiv:0901.3724v1 [gr-qc].
\bibitem{coh}A. G. Cohen, D. B. Kaplan, and A. E. Nelson , Phys. Rev. Lett. \textbf{82}, 4971 (1999).
\bibitem{Li}M. Li, Phys. Lett. B \textbf{603}, 1 (2004).
\bibitem{holo}
Q. G. Huang, and Y. Gong, J. Cosmol. Astropart. Phys. \textbf{08},
006 (2004); B. Wang, Y. Gong, and  E. Abdalla, Phys. Lett. B
\textbf{624}, 141 (2005); K. Enqvist, S. Hannestad, and M. S.
Sloth, J. Cosmol. Astropart. Phys. \textbf{02}, 004 (2005); X.
Zhang, and F. Q. Wu, Phys. Rev. D \textbf{72}, 043524 (2005) ; E.
Elizalde, S. Nojiri, S. D. Odintsov, and P. Wang , Phys. Rev. D
\textbf{71}, 103504 (2005); W. Zimdahl, and D. Pavon,
arXiv:astro-ph/0606555v3; H. Kim, H. W. Lee, and Y. S. Myung,
Phys. Lett. B \textbf{632}, 605 (2006); X. Zhang, Phys. Rev. D
\textbf{74}, 103505 (2006); B. Wang, C. Y. Lin, and E. Abdalla,
Phys. Lett. B \textbf{637}, 357 (2006); X. Zhang, and F. Q. Wu,
 Phys. Rev. D \textbf{76}, 023502 (2007);L. N. Granda, and A. Oliveros, Phys. Lett. B \textbf{669}, 275 (2008);
C. J. Feng, Phys. Lett. B \textbf{663}, 367 (2008); R. Horvat, J.
Cosmol. Astropart. Phys. \textbf{10}, 022 (2008); Y. Gong, and J.
Liu, J. Cosmol. Astropart. Phys. \textbf{09}, 010 (2008); B.
Guberina, R. Horvat, and H. Nikolic , J. Cosmol. Astropart. Phys.
\textbf{01}, 012 (2007); R. G. Cai ,B. Hu, and Y. Zhang ,
arXiv:0812.4504v2 [hep-th]; N. Cruz, P. F. G. Diaz, A. R.
Fernandez, and G. Sanchez, arXiv:0812.4856v1 [gr-qc]; Y. X. Chen,
and Y. Xiao, arXiv:0812.3466v1 [hep-th]; H. Wei, arXiv:0902.2030v1
[gr-qc]; H. M. Sadjadi, and N. Vadood, J. Cosmol. Astropart. Phys.
\textbf{08}, 036 (2008); C. J. Feng, Phys. Lett. B \textbf{672},
94 (2009).
\bibitem{Hsu}S. D. H Hsu, Phys. Lett. B \textbf{594}, 13 (2004).
\bibitem{hua}Q. Huang, and M. Li, J. Cosmol. Astropart. Phys. \textbf{08}, 013 (2004).
\bibitem{bigrip}R. R. Caldwell, M. Kamionkowski, and N. N. Weinberg, Phys. Rev. Lett. \textbf{91}, 071301 (2003);
S. Nojiri, S. D. Odintsov, and S. Tsujikawa, Phys. Rev. D
\textbf{71}, 063004 (2005).
\bibitem{odin}S. Nojiri, and S. D. Odintsov, Gen. Rel. Grav. \textbf{38}, 1285 (2006); E. Elizalde, S. Nojiri, S.
D. Odintsov, and P. Wang, Phys. Rev. D \textbf{71}, 103504 (2005).
\bibitem{new}Y. Z. Ma, Y. Gong, and X. Chen, arXiv: 0901.1215v1 [astro-ph].
\end{thebibliography}
\end{document}